\documentclass[twocolumn,amsmath,floats,showpacs,nofootinbib]{revtex4}
\usepackage[usenames]{color}
\usepackage{graphicx}
\usepackage{epstopdf}
\usepackage{textcomp}
\usepackage{hyperref}
\usepackage{multirow}
\usepackage{tikz}
\usepackage{rotating}
\hypersetup{colorlinks=true}

\begin{document}

\title{Influence of electrical and magnetic fields on the excess current in HTSC - normal metal point contacts}

\author{I.K. Yanson, V.V. Fisun, L.F. Rybaltchenko, N.L. Bobrov, M.A. Obolenskii,
and A.V. Bondarenko}
\affiliation{Institute for Low Temperature Physics and Engineering of the Ukrainian Academy of Sciences, 47 Lenin avenue, 310164 Kharkov, Ukraine\\
Email address: bobrov@ilt.kharkov.ua}

\published {\href{http://fntr.ilt.kharkov.ua/fnt/pdf/18/18-6/f18-0586e.pdf}{Fiz. Nizk. Temp.}, \textbf{18}, 586 (1992)}
\date{\today}

\begin{abstract}
\end{abstract}

\maketitle
\begin{figure}[]
\includegraphics[width=8.5cm,angle=0]{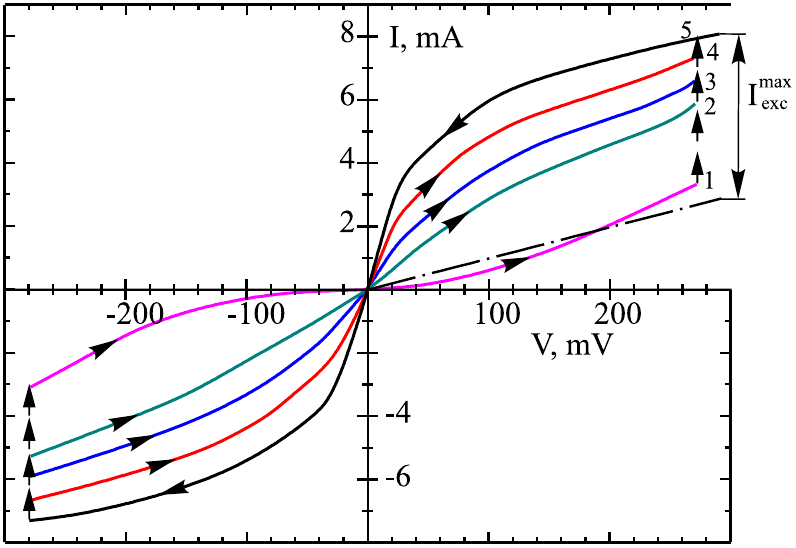}
\caption[]{The effect of I-V characteristic transition between states with different $I_{exc}$ (curves 1-5) YBaCuO-Ag single-crystal point
contact when reaching critical biases. $T=4.2\ K$, $H=0$.}
\label{Fig1}
\end{figure}

It is known that contacts of microscopic sizes with direct conductivity usually called point contacts (PC) can evolve states, which are characterized by high electric fields and current densities. For the first time the S-c-N point contacts made of YBaCuO-Ag showed the effect, which was not observed on traditional superconductors: due to the bias in the point contact the I-V characteristics (IVC) are observed to switch over to the branches with different excess current (Fig.\ref{Fig1}). The IVC of the PC is stable at biases which don't exceed the critical value of about 250~$mV$. At higher biases $I_{exc}$ increases, when the normal metal (Ag) electrode is positive, and decreases as the electrodes change their signs. The IVC switching over to branches with different $I_{exc}$ may be realized with
any discreteness which is dependent on the bias in the contact and the time of PC exposure at a given bias. The dependence obeys the logarithmic law $\delta I_{exc}\sim \text{ln}t$ if the PC bias is several millivolts below the
critical value. On reaching some critical value, the switching over may practically be instantaneous.

\begin{figure}[]
\includegraphics[width=8.5cm,angle=0]{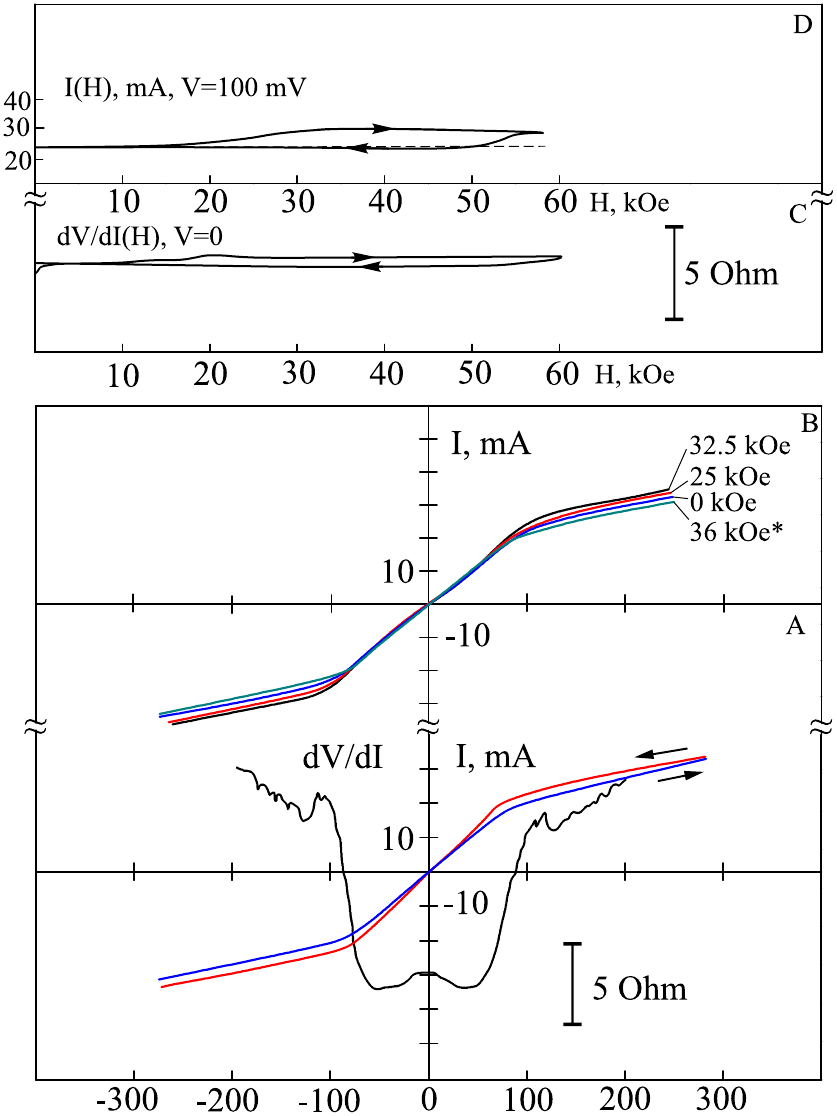}
\caption[]{The magnetic field influence on the magnitude of $I_{exc}$:
IVC and $dV/dI(V)$ in zero magnetic field (a); IVC in different fixed magnetic fields (IVC* corresponds to the point contact in re-assumed magnetic field sweeping down) (b); $dV/dI(H)$ dependence at $V=0$(c); $I_{exc}(H)$ dependence at $V=100\ mV$.}
\label{Fig2}
\end{figure}

\begin{figure}[]
\includegraphics[width=8.5cm,angle=0]{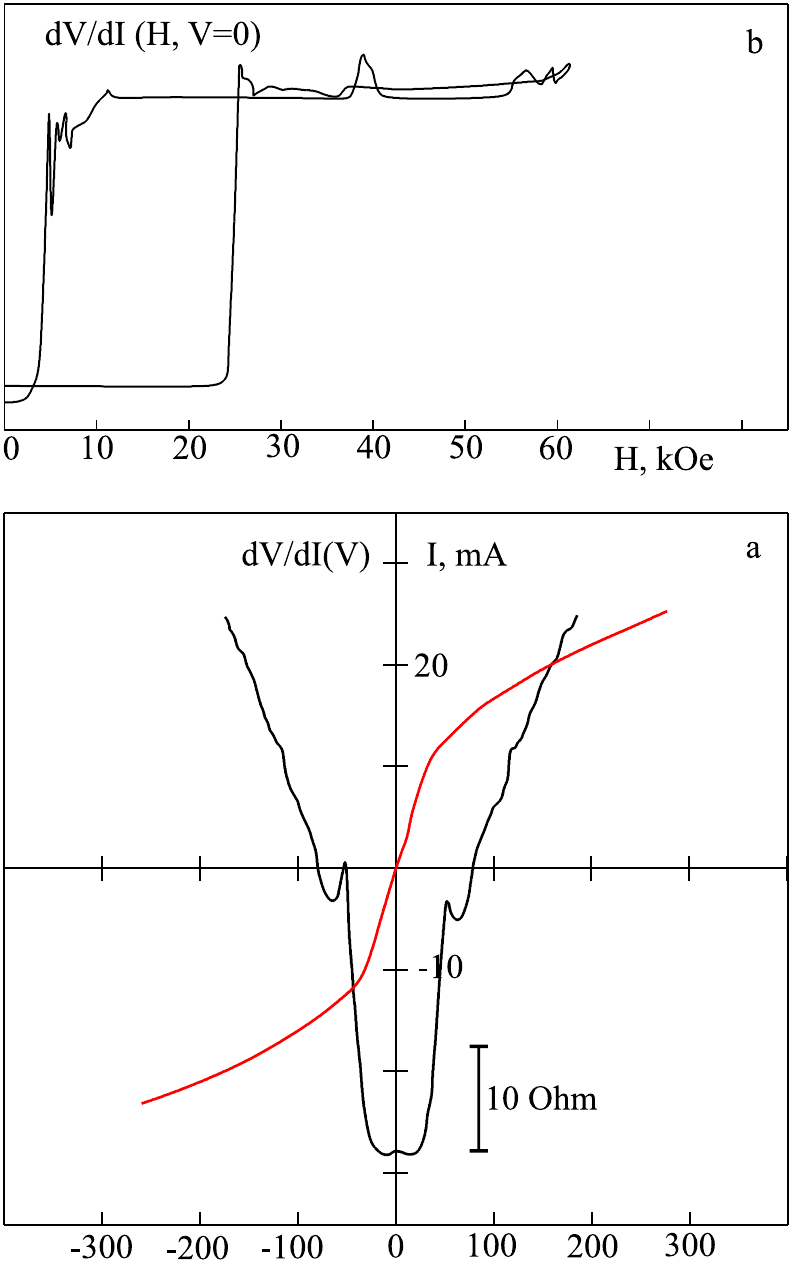}
\caption[]{The magnetic field influence on the magnitude of differential resistance of the point contact at zero voltage: IVC and $dV/dI(V)$ in zero magnetic field (a); $dV/dI(H)$ dependence at $V=0$(b).
}
\label{Fig3}
\end{figure}

The temperature rise to 120~$K$ does not influence the intermediate value of $I_{exc}$; as the temperature returns to its initial point, the chosen intermediate $I_{exc}$ persists.

The nature of the extraordinary IVC behavior may be connected with a change in the HTS electron structure in the near-contact region, i.e. with the bias-caused transition of the electrons from the CuO-1 plane to the CuO-2 one. The holes formed in the CuO-1 plane cause the critical parameters to increase \cite{1}.

We found that some S-c-N PCS displayed an unusual behavior in the magnetic field. The IVC and $dV/dI(V)$ of the PC having the property of the bias-caused IVC switching over to branches with different $I_{exc}$ in the zero magnetic field are shown in Fig.\ref{Fig2},a.
The $I_{exc}$ is seen (Fig.\ref{Fig2},b,c) to increase, as the
magnetic field generated by the superconducting solenoid grows; with a decrease in the external field $I_{exc}$ drops near the turning point, the $I_{exc}$ drop being
sharper than its growth. $I_{exc}$ can even go below its
zero magnetic field value (Fig.\ref{Fig2},b, the IVC marked with*). The zero-bias differential resistance changes slightly too (Fig.\ref{Fig2},c). Several characteristic parts may by separated in the hysteresis curve $I(H)$. In a weak field to about 17~$kOe$ the current is practically invariable, then it starts to increase (10\%) reaching the maximum at $H=40\ kOe$. In the vicinity of the turning point the current is constant in both the branches in the field range of about 7~$kOe$. On removing the magnetic field, the excess current is restored completely.

We also observed another type of the IVC and $dV/dI(V)$ dependences in the external magnetic field. The IVC and $dV/dI(V)$ of PC in the zero magnetic field are shown in Fig.\ref{Fig3},a, and the hysteresis dependence of the differential resistance at $V = 0$ in the magnetic field is illustrated in Fig.\ref{Fig3},b. At some value of the magnetic field the differential resistance increases sharply; as a result, the PC becomes almost dielectric and the $I_{exc}$ is completely suppressed. The
resulting characteristic $R_D(H)$, similary to $I(H)$, displays a hysteretic behavior. As the field is removed, $R_D$ returns to the value close to the initial one.

We believe that the observed hysteresis characteristics $R_D(H)$ and $I_{exc}(H)$ can both be accounted for
by the same factor - the twinning planes present in the studied single crystals and reversible elastic deformation of the HTS single crystal. In fact, the metaloxide high-$T_c$ superconductors are actually heterophase systems with a developed twin structure, which significantly influences the superconducting characteristics of these materials. The presence of twinning planes \cite{2} may lead to a $\sim5\%$ rise of $T_c$.
There are, however, serious grounds for treating the twinning boundaries as nonsuperconducting areas. The lack of unified standpoint concerning the effect of the twinning boundaries on the superconducting properties of HTS materials does not exclude the determining role of the mutual position and electric interaction of copper and oxygen atoms (ions), as well as of the short coherence length in these compounds as compared to that in traditional superconductors. The difference in the extraordinary' behavior between $R_D(H)$ and $I_{exc}(H)$ depends on the location of the
Miming plane in the contact region. For contacts with considerable changes in the PC $R_D(H)$ in the
magnetic field the twinning boundary or other feature of the crystal lattice is probably in the immediate proximity of the N-S boundary of PC. Under the influence of the magnetic field gradient it can move to the contact region and thus increase of the contact resistance since the twin boundaries may be nonsuperconducting \cite{3}.

The increase in the excess current in the magnetic field observed for the contact in Fig.\ref{Fig2} may be due to the compression of the near-contact region at the expense of the difference between the magnetic susceptibilities of the twin boundary and the single crystal, which would lead to higher critical parameters \cite{4}. The change of the magnetic field gradient would eliminate the compression and, consequently, lower the critical parameters. This would lead to $I_{exc}$ a little
lower than the initial value.

There were several attempts to study the interaction of the magnetic and electric field effects upon the PC. The experiment has not revealed the effect of the magnetic field upon the value of the critical electric bias causing the excess current to increase or decrease.


\begin{thebibliography}{}

\bibitem{1} V.I. Kudinov, A.I. Kirilyuk,  N.M. Kreines,  \href{https://doi.org/10.1016/0375-9601(90)90298-3}{Phys. Lett.A} \textbf{151}, 358 (1990).
\bibitem{2} I.N. Khlyustikov and A.I. Buzdin, \href{http://ufn.ru/ufn88/ufn88_5/Russian/r885b.pdf}{Usp. Fiz. Nauk.} \textbf{155}, 47 (1988).
\bibitem{3} M. Sarikaya, R. Kikuchin, I.A. Aksay, \href{https://doi.org/10.1016/0921-4534(88)90010-X}{Physica C} \textbf{152}, 161 (1988).
\bibitem{4} C. W. Chu, P. H. Hor, R. L. Meng, L. Gao, Z. J. Huang, and and Y. Q. Wang, \href{https://doi.org/10.1103/PhysRevLett.58.405}{Phys. Rev. Lett.} \textbf{58}, 405 (1987).



\end{thebibliography}
\end{document}